\documentclass[times,doublespace]{paper}

\usepackage{algorithmicx,algpseudocode,algorithm}

\usepackage{moreverb}
\usepackage[colorlinks,bookmarksopen,bookmarksnumbered,citecolor=red,urlcolor=red]{hyperref}
\usepackage{graphics,graphicx,epic,eepic,amssymb,latexsym,amsmath,amssymb,amsmath,picinpar,epstopdf,times,epsfig}

\usepackage{cite}

\newcommand\BibTeX{{\rmfamily B\kern-.05em \textsc{i\kern-.025em b}\kern-.08em
T\kern-.1667em\lower.7ex\hbox{E}\kern-.125emX}}

\newcommand{\rr}{\mathop{{\rm I}\mskip-4.0mu{\rm R}}\nolimits}

\begin{document}

%\runningheads{G.~Franz\`e \emph{et al.}}{Constrained MPC scheme for NB system with partial state measurements}
\title{HAPD aircraft Uncertain Norm-Bounded Mathematical model}

%\author{G.~Franz\`e \affil{1}, M.~Mattei \affil{2}, L.~Ollio \affil{3} and V.~Scordamaglia \affil{3} \corrauth}
\author{G.~Franz\`e , M.~Mattei, L.~Ollio and V.~Scordamaglia}

% \address{\affilnum{1}DIMES,
%University of Calabria, Rende, 87036, IT \break
% \affilnum{2}DIII, University della Campania {\em Luigi Vanvitelli}, Aversa (CE), 81031, IT\break
%  \affilnum{3}DIIES, University of Reggio Calabria, Reggio Calabria, 89060, IT}

%%%
%%% \corraddr{V. Scordamaglia, DIIES, University of Reggio Calabria, Reggio Calabria, 89060, Italy. E-mail: valerio.scordamaglia@unirc.it}

\maketitle

\begin{abstract}

In this paper an uncertain norm-bounded mathematical model for the UAV High Altitude Performance Demonstrator (HAPD) designed by Italian Aerospace Research Center (CIRA) is carried out. The linear state space description aims to describe the non-linear aircraft dynamic inside the operating envelope characterized by the following bounds  true air speed between $17\, m/s$ and $23\,m/s$ and altitude from $300 \, m$ to $700\, m$.

\end{abstract}

\begin{figure}[ht]
\centering
\epsfig{figure=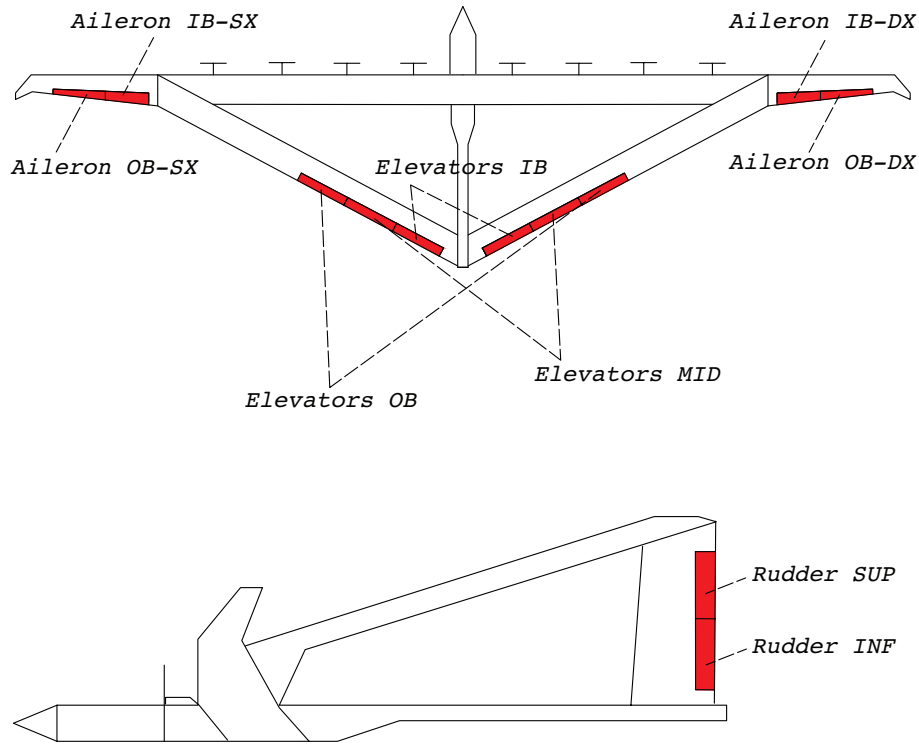,scale=0.5}
%\epsfig{figure=aereo_HPDA,scale=0.6}
  %width=8cm,height=4.5cm}
  \caption{HAPD model: twelve control surfaces and eight available  propellers }\label{HPAD}
\end{figure}

\section{Description of the UAV mathematical model}

The HAPD shown in Fig.\ref{HPAD},  is an over-actuated UAV equipped with the following redundant aerodynamic control surfaces: three pairs of elevators, namely inboard (IB), middle (MID) and outboard (OB), two pairs of ailerons, namely inboard and outboard, and two rudders, namely upper (SUP) and lower (INF). The thrust is generated by eight independent, electrically powered, propellers. In view of  the light weight and the high wing aspect ratio, the HAPD flexibility dynamics have to be taken into account by increasing the number of states with respect to  classical aircraft rigid body mathematical models, see e. g. \cite{StFl2003}. 

As detailed in \cite{CiSo08}, assume the following hypotheses:  the inertia matrix $I$ does not depend on the aircraft elastic deformations; the linear elastic theory can be exploited to model the aero-elastic dynamics;  aero-elastic modes are quasi-stationary.
Starting from these premises, consider  the {\it{polar form}} of the nonlinear equations of the 6DoF motion: 
\begin{equation}\label{motion1}
M \dot{V} = T \cos\alpha \cos \beta-D+
Mg_{1}
\end{equation} 
\begin{equation} V M
\dot{\beta}=-T\cos\alpha \sin\beta+Y -M V
r+Mg_{2}
\end{equation}
\begin{equation}
MV\cos\beta\dot{\alpha}=-T\sin\alpha-L+M V
q+Mg_{3}
\end{equation}
\begin{equation}
I_{x}\dot{p}-I_{xz}\dot{r}=\overline{L}+qr(I_{y}-I_{z})+pqI_{xz}
\end{equation}
\begin{equation}
I_{y}\dot{q}=\overline{M}+rp(I_{z}-I_{x})+(r^{2}-p^{2})I_{xz}
\end{equation}
\begin{equation}\label{motion1bis}
-I_{xz}\dot{p}+I_{z}\dot{r}=\overline{N}+pq(I_{x}-I_{y})-qrI_{xz}
\end{equation}
\begin{equation}
\dot{\phi}=p+q\tan{\theta}\sin\phi+r\tan\theta\cos\phi
\end{equation}
\begin{equation}\label{motion2}
\dot{\theta}=q\cos\phi-r\sin\phi
\end{equation} 
where $T$ is the thrust, $V_{TAS}=\|V_B-V_W\|$ the true air speed,$V_B=(u_B,v_B,w_B)^T$ the 6DoF linear velocity vector, $V_W=(u_W,v_W,w_W)^T$ the atmospheric wind velocity vector, $V=\|V_B\|$, $\omega_B=(p,q,r)^T$ the rotational velocity vector, $\phi$ the roll angle, $\theta$ the pitch angle, $\alpha = \arctan \left (\frac{w_B-w_w}{u_B-u_w} \right )$ the angle of attack,
$\beta=\arcsin \left( \frac{v_B-v_w}{V} \right )$ the sideslip angle,  $I_x$, $I_y$, $I_z$, $I_{xz}$
the moments and products of inertia in body axes, $M$ the aircraft mass and
\begin{equation}
\begin{array}{*{20}{l}}
g_{1} =g(-\cos\alpha\cos\beta\sin\theta+\sin\beta\sin\phi\cos\theta+\sin\alpha\cos\beta\cos\phi\cos\theta)\\
g_{2} =g(\cos\alpha\sin\beta\sin\theta+\cos\beta\sin\phi\cos\theta-\sin\alpha\sin\beta\cos\phi\cos\theta)\\
g_{3} =g(\sin\alpha\sin\theta+\cos\alpha\cos\phi\cos\theta),
\end{array}
\end{equation} 
with $g$ the gravity acceleration. \\
Notice  that a formal definition of  forces and moments involved in (\ref{motion1})-(\ref{motion1bis}) requires the use of  the flexibility dynamics which in turn prescribes the introduction  of additional state variables for  $n_a$ aero-elastic modes, namely the generalized state variables $\eta_i$ and  $\dot{\eta}_i,\,i = 1,\ldots,n_a.$ 
These dynamics are modelled by means of interacting second order linear state space descriptions:
\begin{equation}\label{model4}
M_{\eta_i} \ddot{\eta_i}+\zeta_{\eta_i}\dot{\eta_i} + M_{\eta_i} \omega_{\eta_i} \eta_i=Q_{\eta_i},\,\,i=1,\ldots,n_a,
\end{equation}
where $M_{\eta_i}$ is the generalized mass of the $i-th$ aero-elastic mode, $\zeta_{\eta_i}$   the generalized damping coefficient,  $\omega_{\eta_i}$  the generalized natural  frequency and  $Q_{\eta_i}$ the generalized force. 

\noindent The evolution of the aerodynamic forces and moments  acting on (\ref{motion1})-(\ref{motion1bis}) and (\ref{model4}) depends on both the rigid body and the  flexibility state variables:
\begin{equation}\label{L_eqn}
\begin{array}{ll}
L  = & \!\!\!\!\!\!\frac{\rho V_{TAS}^2 S}{2} \left [  C_L \left ( \!\alpha, \beta,p,q,r,\delta_{sup} \! \right ) \! +\! \displaystyle\sum_{i=1}^{n_a} C_{L_{\eta_i}} \eta_i %\vspace{0.1 cm}
%\\
% & 
+\!\displaystyle\sum_{i=1}^{n_a} C_{L_{\dot{\eta}_i}} \dot{\eta}_i 
\right ] 
\end{array}
 \end{equation}
\begin{equation}\label{D_eqn}
\begin{array}{ll}
D  = & \!\!\!\!\!\!\frac{\rho V_{TAS}^2 S}{2} \left [  C_D \left ( \!\alpha, \beta,p,q,r,\delta_{sup} \! \right ) \! +\! \displaystyle\sum_{i=1}^{n_a} C_{D_{\eta_i}} \eta_i %\\
% &
+ \!\displaystyle\sum_{i=1}^{n_a} C_{D_{\dot{\eta}_i}} \dot{\eta}_i
\right ]
\end{array}
\end{equation}
\begin{equation}\label{Y_eqn}
\begin{array}{ll}	
Y  = & \!\!\!\!\!\!\frac{\rho V_{TAS}^2 S}{2} \left [  C_Y \left ( \!\alpha, \beta,p,q,r,\delta_{sup} \! \right ) \! +\! \displaystyle\sum_{i=1}^{n_a} C_{Y_{\eta_i}} \eta_i %\\
% &
+ \!\displaystyle\sum_{i=1}^{n_a} C_{Y_{\dot{\eta}_i}} \dot{\eta}_i
\right ] 
\end{array}
\end{equation}
\begin{equation}\label{M_eqn}
\begin{array}{ll}
\bar{L}  = & \!\!\!\!\!\!\frac{\rho V_{TAS}^2 S b}{2} \left [  C_l \left ( \!\alpha, \beta,p,q,r,\delta_{sup} \! \right ) \! +\! \displaystyle\sum_{i=1}^{n_a} C_{l_{\eta_i}} \eta_i 
+ \!\displaystyle\sum_{i=1}^{n_a} C_{l_{\dot{\eta}_i}} \dot{\eta}_i
 \right ] 
\end{array}
\end{equation}
\begin{equation}\label{barL_eqn}
\begin{array}{ll}
\bar{M}  = & \!\!\!\!\!\!\frac{\rho V_{TAS}^2 S c}{2} \left [  C_m \left ( \!\alpha, \beta,p,q,r,\delta_{sup} \! \right ) \! +\! \displaystyle\sum_{i=1}^{n_a} C_{m_{\eta_i}} \eta_i +  \!\displaystyle\sum_{i=1}^{n_a} C_{m_{\dot{\eta}_i}} \dot{\eta}_i 
%\\
%&
\right ] 
\end{array}
 \end{equation}
\begin{equation}\label{barN_eqn}
\begin{array}{ll}
\bar{N}  = & \!\!\!\!\!\!\frac{\rho V_{TAS}^2 S b}{2} \left [ C_n \left ( \!\alpha, \beta,p,q,r,\delta_{sup} \! \right ) \! +\! \displaystyle\sum_{i=1}^{n_a} C_{n_{\eta_i}} \eta_i
%\\
%&
+ \!\displaystyle\sum_{i=1}^{n_a} C_{n_{\dot{\eta}_i}} \dot{\eta}_i
 \right ] 
\end{array}
\end{equation}
\begin{equation}\label{barQ_eqn}
\begin{array}{ll}
\bar{Q}_{\eta_i}  = & \!\!\!\!\!\!\frac{\rho V_{TAS}^2 S}{2} \left [ C_0^i+C_{\alpha}^i \alpha + C_{\beta}^i \beta + C_p^i p + C_q^i q+ C_r^i r  
+ C_{\delta_{sup}}^i \delta_{sup}
\right . \\
 & \left . + 
 \displaystyle \sum_{j=1}^{n_a}C^{ij}_{\eta_j} \eta_j +  \displaystyle \sum_{j=1}^nC^{ij}_{\dot{\eta}_j} \dot{\eta}_j\!\right ] 
  \!
\end{array} 
\end{equation}
where the numerical values of the main involved variables  are reported in Table \ref{HPDA-coeff}. Moreover,  $\delta_{sup}$  is a vector accounting for  control surfaces deflections, $\rho$ the air density at the flying altitude and all the terms indexed by $C$
%, accompained by some pedices and/or apices, 
refer to  adimensional aerodynamic coefficients arising from  the common hypothesis that aerodynamic forces are affine functions of the motion  variables and  generalized states $\eta_i$ and $\dot{\eta}_i.$\\
Note also that the flexible UAV mathematical model requires the calculation of both generalized masses, damping coefficiens, natural frequencies  and aerodynamic coefficients. The first   are computed via finite structural elements methods, while aerodynamic coefficients are obtained  by resorting to CFD (Computational Fluid Dynamics) calculations \cite{WaSch88} and/or wind tunnel or flight tests.

\begin{table}
\caption{HAPD Main Parameters} \label{HPDA-coeff}
\begin{center}
\begin{tabular}{|c||c|c|}
\hline
{\bf{\normalsize{Names}}} & {\bf{\normalsize{Values}}} & {\bf{\normalsize{Units}}}\\ \hline
\normalsize{Wing Area (S)} & \normalsize{13.5} & \normalsize{$m^2$}\\ \hline
\normalsize{Wing Span ($b$)} & \normalsize{16.55} & \normalsize{$m$}\\ \hline
\normalsize{Mean Chord ($c$)} & \normalsize{0.557} & \normalsize{$m$} \\ \hline
\normalsize{Mass (M)} & \normalsize{184.4} & \normalsize{$kg$} \\ \hline
\normalsize{Moment of Inertia $I_x$} & \normalsize{$1.997\cdot 10^3$} & \normalsize{$kg \cdot m^2$} \\ \hline
\normalsize{Moment of Inertia $I_y$} & \normalsize{$258.6$} & \normalsize{$kg \cdot  m^2$} \\ \hline
\normalsize{Moment of Inertia $I_z$} & \normalsize{$2.196 \cdot 10^3$} & \normalsize{$kg \cdot m^2$} \\ \hline
\normalsize{Product of Inertia $I_{xz}$} & \normalsize{$-66.3$} & \normalsize{$kg \cdot m^2$} \\ \hline
%\normalsize{Actuators Undamped Natural Frequency ($f_a$)} & \normalsize{$13.35$} & \normalsize{$Hz$} \\ \hline
%\normalsize{Actuators Damping Ratio ($\zeta_a$) } & \normalsize{$1$} & \normalsize{$-$} \\ \hline
%\normalsize{Sensors Bandwidth ($f_s$)} & \normalsize{$13.35$} & \normalsize{$Hz$} \\ 
%\hline Outboard Elevators Slew Rate & $\pm 200$  & $deg/s$ \\ \hline
\normalsize{Ailerons Slew Rates} & \normalsize{$\pm 200$} & \normalsize{$deg/s$} \\ \hline
 \normalsize{Elevators Slew Rates} & \normalsize{$\pm 200$} & \normalsize{$deg/s$} \\  \hline
\normalsize{Rudders Slew Rates} & \normalsize{$\pm 200$} & \normalsize{$deg/s$} \\ \hline
\normalsize{Ailerons deflections}  & \normalsize{$\pm 25$} & \normalsize{$deg$} \\ \hline
\normalsize{Elevators deflections} & \normalsize{$\pm 25$} & \normalsize{$deg$} \\ \hline
\normalsize{Rudders deflections} & \normalsize{$\pm 25$} & \normalsize{$deg$} \\ \hline
\end{tabular}
\end{center}
\end{table}

\section{Uncertain Norm-Bounded modeling}
%\note{R2-13}

By considering only the two slower and most significant aero-elastic modes accounting for  symmetrical ($\eta_1=\eta_s$)
and  asymmetrical ($\eta_2=\eta_a$) 
aircraft deformations, faster aero-elastic dynamics being considered instantaneous, the high nonlinear model description 
(\ref{motion1})-(\ref{barQ_eqn})
 has been recast into the following uncertain linear state space description with
uncertainties (or perturbations) appearing in {\color{black} a} feedback loop 
\begin{equation}\label{sys-nb}
  \left\{ \begin{array}{lcl}
    x(t+1) & = & \Phi \, x(t) + G \, u(t) + B_p \, p(t) \\
    y(t) & = & C \, x(t) \\
    q(t) & = & C_q \, x(t) + D_q \, u(t) \\
    p(t) & = & \Delta(t) \, q(t)
  \end{array} \right.
\end{equation}
where $x \in \rr^{12}$ denoting the state, $u \in \rr^{12}$ the
control input, $y \in \rr^{8}$ the output.

The input and state vectors definition  are below reported
$$
x(t)=[V,
\alpha,
\beta,
p,
q,
r,
\phi,
\theta,
\eta_s,
\dot{\eta}_s,
\eta_a,
\dot{\eta}_a
 \!]^T,\,\, u(t)=[\delta_{sup}\,\,\, T]^T
$$
$$ 
\delta_{sup}=\left[
\begin{array}{l} 
 Elevator IB-DX\\
Elevator IB-SX\\
Elevator MID-DX\\
Elevator MID-SX\\
Elevator OB-DX\\
Elevator OB-SX\\
Aileron IB-DX\\
Aileron IB-SX\\
Aileron OB-DX\\
Aileron OB-SX\\
Rudder SUP\\
Rudder INF
\end{array} \right ]
$$ 
$$
y(t)=[V,
\alpha,
\beta,
p,
q,
r,
\phi,
\theta]^T
$$
Moreover, $p , q \in \rr^{12}$
are the additional variables accounting for the uncertainty. %\note{R2-2}
The uncertain operator $\Delta$ may represent either a memoryless, possibly time-varying, matrix with
$\left\| \Delta(t) \right\|_2 = \bar{\sigma}\left( \Delta(t) \right) \le
1$ $\forall t \ge 0$, or a convolution operator with norm, induced by the
truncated $\ell_2$-norm, less than 1 viz.
%
%\begin{equation}\labelwrite{conv}
$$
  \sum_{j=0}^t p(j)^T \, p(j) \le \sum_{j=0}^t q(j)^T \,
  q(j)\, , \forall t \ge 0
$$
%\end{equation}
%
For a more extensive discussion about this type of uncertainty see
\cite{Boyd94}.
Such a representation (\ref{sys-nb})
\normalsize
can be obtained by exploiting the fact that
%a set of linearized  models well approximates aircraft dynamics in wide regions of the flight envelope. Therefore,  
a suitable collection of linear models well approximates aircraft dynamics in wide regions of the flight envelope 
%can be used to describe aircraft dynamics all over this operating envelope 
including steady state and transient flight conditions.
Thus, a Polytopic Linear Differential Inclusion (PLDI) of the HAPD nonlinear model (\ref{motion1})-(\ref{model4}) can be obtained by deriving a convex outer approximation of regions covered by a set of $30$ linearized models around different operating flight conditions characterized by the following bounds on speed and altitude:
 true air speed between $17$ and $23$ \emph{m/s} and altitude from $300$ \emph{m} to $700$ \emph{m}. 
% \note{R1-8}
%%% \begin{color}{black} 
%%%In fact, it is common practice to schedule flight controllers with speed or Mach number and altitude, which has an impact on the air density, because these two variables have an impact on natural frequencies and damping of both the longitudinal and lateral directional flight modes \cite{StFl2003}.\end{color}\\
%
PLDI is then approximated as a Norm-bound Linear Differential Inclusion (NDLI) by exploting the optimization procedure described in \cite{Boyd94}.

\bibliography{mybib}

\end{document}